# FULLERENES AND PROTO-FULLERENES IN INTERSTELLAR CARBON DUST

W. W. Duley & Anming Hu

Department of Physics and Astronomy,

University of Waterloo,

Waterloo, Ontario,

Canada, N2L 3G1

Email: wwduley@uwaterloo.ca

ABSTRACT

Laboratory spectra of hydrogenated amorphous carbon (HAC) particles prepared under a variety of conditions show spectral features at 7.05, 8.5, 17.4 and 18.9 μm (1418, 1176, 575 & 529 cm$^{-1}$) that have been associated with emission from $C_{60}$ molecules. *These lines occur in the spectra even though $C_{60}$ molecules as such are not present in our samples.* It appears that these four spectral lines in HAC can instead be associated with precursor molecules or "proto-fullerenes" that subsequently react to yield $C_{60}$. We develop a model tracing the evolution and de-hydrogenation of HAC dust and show that the observation of an emission feature at 16.4 μm (610 cm$^{-1}$) in astronomical spectra signals the presence of the pentagonal carbon rings required for the formation of fullerenes. We suggest that the set of four IR emission lines previously identified with $C_{60}$ in many objects that also show the 16.4 μm feature and other polycyclic aromatic hydrocarbon bands arise from proto-fullerenes rather than $C_{60}$. Tc1 is an example of a source in which de-hydrogenation has proceeded to the point where only fullerenes are present.

Subject headings: infrared: ISM – ISM: lines and bands – ISM: molecules

1. INTRODUCTION

Since the detection of $C_{60}$ and $C_{70}$ in Tc1, a young planetary nebula (Cami et al. 2010), similar spectral features have been found in a number of other planetary nebulae (PNe) (Garcia-Hernandez et al. 2010, 2011b,c) as well as in the proto-planetary nebula IRAS 01005+7910 (Zhang & Kwok 2011). A number of these features have also been found in spectra of the reflection nebulae (RNe) NGC 7023 and NGC 2023 (Sellgren et al. 2010) and they appear in the emission spectrum of DY Cen, a hydrogen-deficient R Coronae Borealis (RCB) star (Garcia-Hernandez et al. 2011a), but not in other RCB stars. In all cases, with the exception of Tc1, the emission lines associated with $C_{60}$ are accompanied by emission from polycyclic aromatic hydrocarbons (PAHs) and other hydrocarbon species.

The emission spectrum of $C_{60}$ is characterized by IR lines at 7.05, 8.5, 17.4 and 18.9 μm corresponding to the four $T_{1u}$ vibrational modes of this highly symmetric molecule (Cami et al. 2010). Analysis of the relative intensities of these emission lines, together with excitation conditions and transition energies in Tc1, indicates that $C_{60}$ emitters in this source may be components of dust grains rather than gas-phase molecules and line ratios for the four IR transitions imply that the emitters are in Boltzmann equilibrium at a temperature $T_{ex} \approx 330$ K. PAH and aliphatic hydrocarbon molecules are completely absent in Tc1 (Cami et al. 2010) leading to the suggestion that fullerenes, and $C_{60}$ in particular, may be the products of radiative or thermal processing of hydrogenated amorphous carbon (HAC or a:C-H) dust (Cami et al. 2010). Previous studies have shown that large carbon molecules including $C_{60}$ are released in the photo-chemical processing of HAC (Scott et al. 1997).

In this paper we examine a possible connection between chemistry in HAC grains and the appearance of IR spectral features characteristic of $C_{60}$. This is a continuation of our earlier study (Hu & Duley 2008) that noted a remarkable similarity between 16-20 μm spectra of small HAC particles and those seen from various locations in the reflection nebula NGC 7023 (Sellgren et al. 2007). This earlier work is now re-examined in the context of the possible detection of $C_{60}$ in this and other nebulae as well as in a variety of other objects. We find that IR emission spectra of carbonaceous dust present in different circumstellar and interstellar environments can be used to trace the evolution of hydrogen-rich, primarily aliphatic material, to proto-fullerenes and fullerene-rich amorphous carbons.

## 2. RESULTS AND ANALYSIS

Thin film deposits of amorphous carbon (a:C) and HAC, assembled from carbon nanoparticles, were prepared by the laser ablation of graphite as reported earlier (Hu, Alkhesho & Duley 2006, Hu & Duley 2008). To evaluate the effect of ambient conditions, samples were deposited in the presence of $H_2$, He, acetylene ($C_2H_2$) and ethylene ($C_2H_4$) gases as well as in vacuum. Deposition in vacuum produces particles that are predominantly a:C, but all deposits contain some hydrogen as this is always present in the parent graphite: these conditions are designated as "vacuum (H)". These samples have the structure shown in Figure 1 of Hu & Duley 2008. All spectra were recorded at 300K which is similar to the excitation temperature reported for $C_{60}$ in Tc1 (Cami et al. 2010). A small shift to lower energy is expected at higher temperatures. A thin film of $C_{60}$ was also prepared in vacuum using the same laser vaporization and deposition technique. IR spectra were obtained in the surface enhanced Raman (SERS) format as this is known to yield accurate molecular frequencies for both Raman and IR transitions (Moskovits, DiLella & Maynard 1988, Matejka et al. 1996).

Figures 1 & 2 show a comparison between SERS spectra of thin cluster-assembled carbon films deposited in vacuum (H), He, $H_2$, $C_2H_2$ and $C_2H_4$ with that of $C_{60}$ in the region of the $T_{1u}(1)$, $T_{1u}(2)$, $T_{1u}(3)$ and $T_{1u}(4)$ lines. The spectrum of $C_{60}$ in this energy range is characterized by intense $A_g(1)$ and $A_g(2)$ transitions at 491 cm$^{-1}$ and 1461 cm$^{-1}$, respectively, associated with breathing modes of the complete $C_{60}$ shell (Ikeda & Uosaki 2008). There is no compelling evidence for these features in spectra of the laboratory samples. A feature near 491 cm$^{-1}$ is marginally present in the vacuum (H) spectrum and appears with low intensity in the $H_2$, He and $C_2H_2$ samples, but this does not appear in that of the sample deposited in $C_2H_4$. A band near the $C_{60}$ $A_{1g}(2)$ transition at 1461 cm$^{-1}$ is observed in the He and $C_2H_2$ samples and is marginally detected in the spectrum of the vacuum (H) film, but it is not found in spectra of films deposited in $H_2$ or $C_2H_4$. Although the relative intensity of the $A_{1g}(1)$ and $A_{1g}(2)$ lines in the Raman spectrum of $C_{60}$ has been found to vary under SERS excitation (Luo et al. 2010), we consider that the absence of strong features at 491 cm$^{-1}$ and 1461 cm$^{-1}$ in Figures 1 & 2 is an indication that $C_{60}$, in the form of individual molecules, is not abundant in our samples. If $C_{60}$ was present, it would have to be bonded to other chemical groups in order to destroy the overall symmetry of the $C_{60}$ cage and suppress the $A_{1g}$ vibrational modes.

Energies of the $T_{1u}(1)$, $T_{1u}(2)$, $T_{1u}(3)$ and $T_{1u}(4)$ lines of $C_{60}$ as detected in Tc1 (Cami et al. 2010) are plotted in Figures 1 & 2 for comparison with laboratory spectra. Spectral features are present at, (or close to), these energies in all five laboratory samples with the exception of the $C_2H_2$ sample where the $T_{1u}(2)$ line is absent. In particular, the $T_{1u}(3)$ and $T_{1u}(4)$ lines occurring at 1177 and 1423 cm$^{-1}$ (8.50 and 7.03μm), respectively, in the spectrum of Tc1 have counterparts in most of the laboratory samples. It is interesting that there is some dispersion in the energy of the $T_{1u}(3)$ feature in the laboratory spectra that may reflect variations seen in astronomical spectra (Table 1). It should be noted at this point that the IR active $T_{1u}$ features of $C_{60}$ are not observed in the Raman spectrum in the absence of the SERS effect (Ikeda & Uosaki 2008, Luo et al. 2010). Their appearance in our SERS spectra would occur if a) $C_{60}$ molecules are indeed present in our samples and that the SERS effect is enhancing the strength of $T_{1u}$ relative to $A_{1g}$ transitions, or b) that the observed features arise from transitions in molecules having lower symmetry and closely related to, but not identical to $C_{60}$. Given the weakness of the normally intense $A_{1g}$ features in our spectra, and guided by the work of Luo et al. (2010) on the effects of SERS enhancement on the spectrum of $C_{60}$, we favor the latter possibility; namely that chemical species similar to $C_{60}$, but not exhibiting the high symmetry associated with the perfect icosohedral structure of this molecule, are present in our laboratory samples. We will refer to these structures as "proto-fullerenes" (PFs). They co-exist with PAH and aliphatic hydrocarbons in HAC and have been seen in other laboratory simulations (Jäger et al. 2008).

Table 1 summarizes the energies of the $T_{1u}$ transitions of $C_{60}$ as detected in Tc1 together with those of emission features in other sources and compares these with spectral features found in our samples. Differences in the energies of the laboratory transitions in these spectra are real and reflect changes in composition between individual samples. They suggest that the chemical route to $C_{60}$ and other fullerenes depends on the initial composition of HAC nano-particles. Slight differences in the energies of astronomical features (table 1) could then arise from changes in the composition of PFs in different astronomical environments, or changes in temperature (Iglesias-Groth et al. 2011). A number of emission lines attributable to $C_{70}$ have also been observed in Tc1 (Cami et al. 2010) and are listed in Table 2, for comparison with laboratory data. Several of the IR active modes in $C_{70}$ occur at energies that result in blending of these features with the $T_{1u}$ modes of $C_{60}$ (Cami et al. 2010). Spectral features at 534, 564 and 578 cm$^{-1}$ (18.7, 17.7 and 17.3 μm) associated with $C_{70}$ in Tc1 are clearly visible in the spectrum of the He sample (Figure 1).

A notable feature of $C_{60}$ and other fullerenes is the incorporation of pentagonal rings in a network of hexagonal carbon rings (Kroto et al. 1985). All fullerenes have 12 pentagonal rings together with a varying number of hexagonal rings but the $C_{60}$ molecule is highly stable because it is the smallest fullerene containing isolated pentagonal rings. The smallest fullerene is the much less stable $C_{20}$ molecule with a cage structure consisting of 12 adjacent pentagonal rings (Prinzbach et al. 2006). The chemical precursors to $C_{60}$ are therefore molecular structures that contain pentagonal rings bonded in other configurations. The presence of these precursors in our samples is indicated by the appearance of characteristic IR bands of pentagonal carbon rings near 1430 and 610 cm$^{-1}$ (7.0 and 16.4 μm) (Moutou et al. 2000). The 610 cm$^{-1}$ (16.4 μm) band is most apparent in the vacuum (H), He and $C_2H_4$ samples (Figures 1 & 2). It is not seen in the $H_2$ and $C_2H_2$ samples. Significantly, $C_{60}$ does not have IR or Raman transitions near 610 cm$^{-1}$ (Mendenez & Page 2000) because both transitions are forbidden in the icosohedral symmetry group. Observation of a 16.4 μm band in astronomical and laboratory spectra is then a clear indication that the chemical precursors to fullerene molecules are present in this material. As the 16.4 μm feature is widely detected in emission spectra of astronomical sources (Smith et al. 2007), it is evident that the conditions for the formation of fullerenes exist in many astronomical environments.

Our laboratory spectra show that samples containing fullerene precursors, but not $C_{60}$, can also have spectral features at 7.05, 8.5, 17.4 and 18.9 μm (1418, 1176, 575 & 529 cm$^{-1}$) (Tables 1-2). This suggests that emission features seen at these wavelengths in many sources (Garcia-Hernandez et al. 2010, 2011a, b, Sellgren et al. 2010, Zhang & Kwok 2011) are attributable to molecules that may lead to the formation of fullerenes, but are not fullerenes as such. This would be consistent with the co-existence of PAH and aliphatic compounds together with PFs in these sources. For example, PAH is found along with PF emission in IRAS 01005+7910 (Zhang & Kwok 2011) and in NGC 7023 (Sellgren et al. 2010) but PAHs are not seen in Tc1 where the emission can be attributed solely to $C_{60}$ and $C_{70}$ (Cami et al. 2010). This indicates that rather special conditions may be required to convert PFs to $C_{60}$ and other fullerenes and that these conditions have not been present in objects such as NGC 7023. Fulleranes (hydrogenated fullerenes) (Cataldo et al. 2010) may be by-products of this conversion, but are not considered here.

We suggest that the evolutionary sequence that converts hydrogen-rich carbons to fullerenes (represented as HAC (a:C-H) → a:C → $C_{60}$ + other fullerenes) can be traced from astrophysical spectra as follows. Stage 1: a reduction in the hydrogen content in HAC as measured by the shift in the "6.2 μm" band from ~ 6.3 to 6.2 μm (~1580 – 1610 $cm^{-1}$) (Pino et al. 2008). Pino et al. (2008) have shown that this shift signals the conversion of aliphatic ($sp^3$) to aromatic ($sp^2$) bonding. Stage 2: a decrease in the ratio I (11.3 μm) / I (6.2 μm) as aromatic rings cluster to form larger ring structures and PFs. Stage 3: conversion of PFs to closed cage structures (eg. $C_{60}$). The 16.4 μm band associated with pentagonal rings would be present in stages 1 & 2, but would be weak or absent in stage 3 material. Tc1 then represents stage 3 in the evolutionary process, while IRAS 01005+7910 (Zhang & Kwok 2011) is an example of stage 2 material that is evolving into stage 3. A similar sequence involving reactions between precursor carbon ring structures to form $C_{60}$ and other fullerenes has been discussed by Wakabayashi & Achiba (1992) and Osterodt et al. (1996).

9Overall, the dehydrogenation of HAC to form PFs and then fullerenes is an endothermic process although the formation of $C_{60}$ and other fullerenes from small carbon clusters is exothermic once an activation energy has been supplied (Suzuki et al. 2001). Since $C_{60}$ and larger fullerenes contain many carbon atoms, the implication is that, unless fullerenes pre-exist in the gas phase, they must form in carbon dust grains, as only larger particles can accommodate the number of atoms required. These particles will be too large to undergo significant temperature spiking by absorbing photons from the interstellar radiation field, but they may transiently reach temperatures in excess of 1000K as a result of the release of stored chemical energy (Duley & Williams 2011). This would be consistent with observations showing IR emission lines from PFs together with those arising from PAHs and aliphatic components (Garcia-Hernandez et al. 2010, 2011a, b). We note that fullerenes, including $C_{60}$, have been detected in the mass spectrum of molecules ejected from HAC following laser heating (Scott et al. 1997) even when $C_{60}$ as such is not present in the parent material. This is consistent with the thermally-induced conversion of PFs to fullerenes during the decomposition process and would represent the end-point of the de-hydrogenation of HAC.

## 3. CONCLUSIONS

Spectral lines at 7.05, 8.5, 17.4 and 18.9 μm (1418, 1176, 575 & 529 cm$^{-1}$) normally associated with the $C_{60}$ molecule are found in HAC nano-particle samples under conditions where $C_{60}$ is not present. These features arise from proto-fullerenes (PFs) and are often accompanied by an emission band at 16.4 μm that is typical of pentagonal carbon rings. We suggest that the $C_{60}$ molecule as such has only been detected in Tc1 and that similar spectra obtained of PPne, PNe,

RNe and other sources are likely indicating the presence of the chemical precursors of $C_{60}$ rather than the icosohedral molecule. Although we cannot confirm that no other fullerenes are to be found in HAC nano-particles, our experimental data indicates that $C_{60}$ is not a significant component. It is then unlikely that other fullerenes will be present under the conditions of our laboratory simulations.

This research was supported by a grant from the NSERC of Canada.

REFERENCES

Cami, J., Bernard-Salas, J., Peeters, E., & Malek, S. E., 2010, Science, 329, 1180

Cataldo, F., Iglesias-Groth, S., & Manchado, A., 2010, in *Fulleranes: The Hydrogenated Fullerenes*, eds. F. Cataldo & S. Iglesias-Groth, Springer, New York, p203

Duley, W. W., & Williams, D. A., 2011, ApJ, 737, L44

García-Hernández, D. A. Manchado, A. García-Lario, P. Stanghellini, L. Villaver, E. Shaw, R. A. Szczerba, R., & Perea-Calderón, J. V., 2010, ApJ, 724, L39

García-Hernández, D. A. Kameswara Rao, N. & Lambert, D. L. 2011a, ApJ, 729, 126

García-Hernández, D. A. Iglesias-Groth, S. Acosta-Pulido, J. A., Manchado, A. García-Lario, P. Stanghellini, L. Villaver, E. Shaw, R. A. & Cataldo, F., 2011b, ApJ, 737, L30

Hu, A., Alkhesho, I., & Duley, W. W., 2006, ApJ, 653, L157

Hu, A., & Duley, W. W., 2008, ApJ, 672, L81

Iglesias-Groth, S., Cataldo, F. & Manchado, A., 2011, MNRAS, 412, 213

Ikeda, K., & Uosaki, K., 2008, J. Phys. Chem. A112, 790

Jäger, C., Mutschke, H., Henning, Th., & Huisken, F., 2008, ApJ, 689, 249

Matejka, P., Stavek, J., Volka, K., Schrader, B., 1996, Appl. Spectros., 50, 409

Menendez, J., & Page, J. B., 2000, in Light Scattering in Solids VIII: Fullerenes, 9 Semiconductor Surfaces, Coherent Phonons, ed. M. Cardona & G. Guntherodt (Berlin: Springer), 27

Moskovits, M., DiLella, D. P., & Maynard, K. J., 1988, Langmuir, 4, 67

Moutou, C. Verstraete, L. Léger, A. Sellgren, K., & Schmidt, W., 2000, A&A, 354, L17

Osterodt, J., Zett, A., & Vogtle, F., 1996, Tetrahedron, 52, 4949

Pino, T., Dartois, E. Cao, A.-T., Carpentier, Y., Chamaillé, Th., Vasquez, R., Jones, A. P., D'Hendecourt, L., & Bréchignac, Ph., 2008, A&A, 490, 665

Prinzbach, H., et al., 2006, Chem. Eur. J., 12, 6268

Scott, A., Duley, W. W., & Pinho, G., 1997, ApJ, 489, L193

Sellgren, K., Uchida, K. I., & Werner, M., 2007, ApJ, 659, 1338

Sellgren, K., Werner, M., Ingalls, J. G., Smith, J. D. T., Carleton, T. M., & Joblin, C., 2010, ApJ, 722, L54

Smith, J. D. T., et al., 2007, ApJ, 656, 770

Suzuki, S., et al., 2001, Euro Phys. J. D, 16, 369

Wakabayashi, T., & Achiba, Y., 1992, Chem. Phys. Lett., 190, 465

Zhang, Y., & Kwok, S., 2011, ApJ, 730, 126

Table 1. Energies (cm$^{-1}$) of T$_{1u}$ spectral lines of C$_{60}$ observed in Tc1 together with similar features seen in other astronomical sources (Cami et al. 2010, Garcia-Hernandez et al. 2010, 2011a, Zhang & Kwok 2010) compared with those of spectral features in HAC laboratory samples measured at 300K. We note that the excitation temperature of C$_{60}$ in Tc1 is ≈ 330K (Cami et al. 2010) and excitation temperatures between 330 and 680K are indicated in other sources (Garcia-Hernandez et al. 2010). Laboratory energies are accurate to ± 0.3 cm$^{-1}$.

| Deposition condition | Energy (cm$^{-1}$) | | | |
| --- | --- | --- | --- | --- |
|  | T$_{1u}$(1) | T$_{1u}$(2) | T$_{1u}$(3) | T$_{1u}$(4) |
| Vacuum (H) | 532 | 575 | 1183w | 1417w |
| He | 522 | 568 | 1183 | 1429 |
| H$_2$ | 527 | 563 | 1163 | 1422 |
| C$_2$H$_2$ | 528 |  | 1163 | 1426 |
| C$_2$H$_4$ | 530 | 571 | 1180 | 1422 |
| Source |  |  |  |  |
| Tc1 | 529 | 575 | 1177 | 1423 |
| M1-20 | 527 | 576 | 1166 | 1425 |
| M1-12 | 528 | 576 | 1166 | 1431 |
| K3-54 | 530 | 575 | 1176 | 1422 |
| SMC 16 | 528.5 | 574 | 1168 | 1425 |
| DY Cen | 532 | 575 |  | 1430 |
| IRAS01005+7910 | 529 | 575 |  | 1420 |

Table 2. Energies (cm$^{-1}$) of spectral lines of C$_{70}$ observed in Tc1 (Cami et al. 2010) compared with those of spectral features in laboratory samples. Laboratory energies are accurate to ± 0.3 cm$^{-1}$.

| Deposition condition | | | Energy (cm$^{-1}$) | | | | |
|---|---|---|---|---|---|---|---|
| Vacuum (H) | 459 | 646 | 677w | | 1140 | 1417w | |
| He | 463w | 644 | | 791 | 1142 | 1420 | 1429 |
| H$_2$ | 473w | | | | 1140 | 1403 | 1422 |
| C$_2$H$_2$ | 464 | 650w | 670 | 796w | | | 1426 |
| C$_2$H$_4$ | 462w | 648 | 671 | 791 | 1133 | 1422 | |
| | | | | | | | |
| Source | | | | | | | |
| Tc1 | 458 | 642 | 676 | 799 | 1133 | 1414 | 1430 |

Figure captions

Figure 1. SERS spectra at low energy recorded for HAC nano-particle samples deposited by femtosecond ablation of graphite in vacuum, He, $H_2$, $C_2H_2$ and $C_2H_4$. The bottom spectrum is that of a thin film deposited from the laser ablation of $C_{60}$ in vacuum. The strong transition of $C_{60}$ at 491 cm$^{-1}$ is not evident in the laboratory samples indicating that $C_{60}$ molecules as such (ie. having full icosohedral symmetry) are not components of the deposited films. Dashed lines indicate the energies of the $T_{1u}(1)$ and $T_{1u}(2)$ transitions of $C_{60}$ measured by Cami et al. (2010) in the spectrum of Tc1. Note that a number of the samples have spectral features at these energies that can be associated with proto-fullerenes. A feature at 610 cm$^{-1}$ (16.4 μm) in some samples signals the presence of pentagonal rings.

Figure 2. SERS spectra at higher energy recorded for HAC nano-particle samples deposited by femtosecond ablation of graphite in vacuum, He, $H_2$, $C_2H_2$ and $C_2H_4$. The bottom spectrum is that of a thin film deposited from the laser ablation of $C_{60}$ in vacuum. The strong transition of $C_{60}$ at 1461 cm$^{-1}$ is not evident in the laboratory samples indicating that $C_{60}$ molecules as such (ie. having full icosohedral symmetry) are not components of the deposited films. Dashed lines indicate the energies of the $T_{1u}(3)$ and $T_{1u}(4)$ transitions of $C_{60}$ measured by Cami et al. (2010) in the spectrum of Tc1. Note that a number of the samples have spectral features at these energies that can be associated with proto-fullerenes.

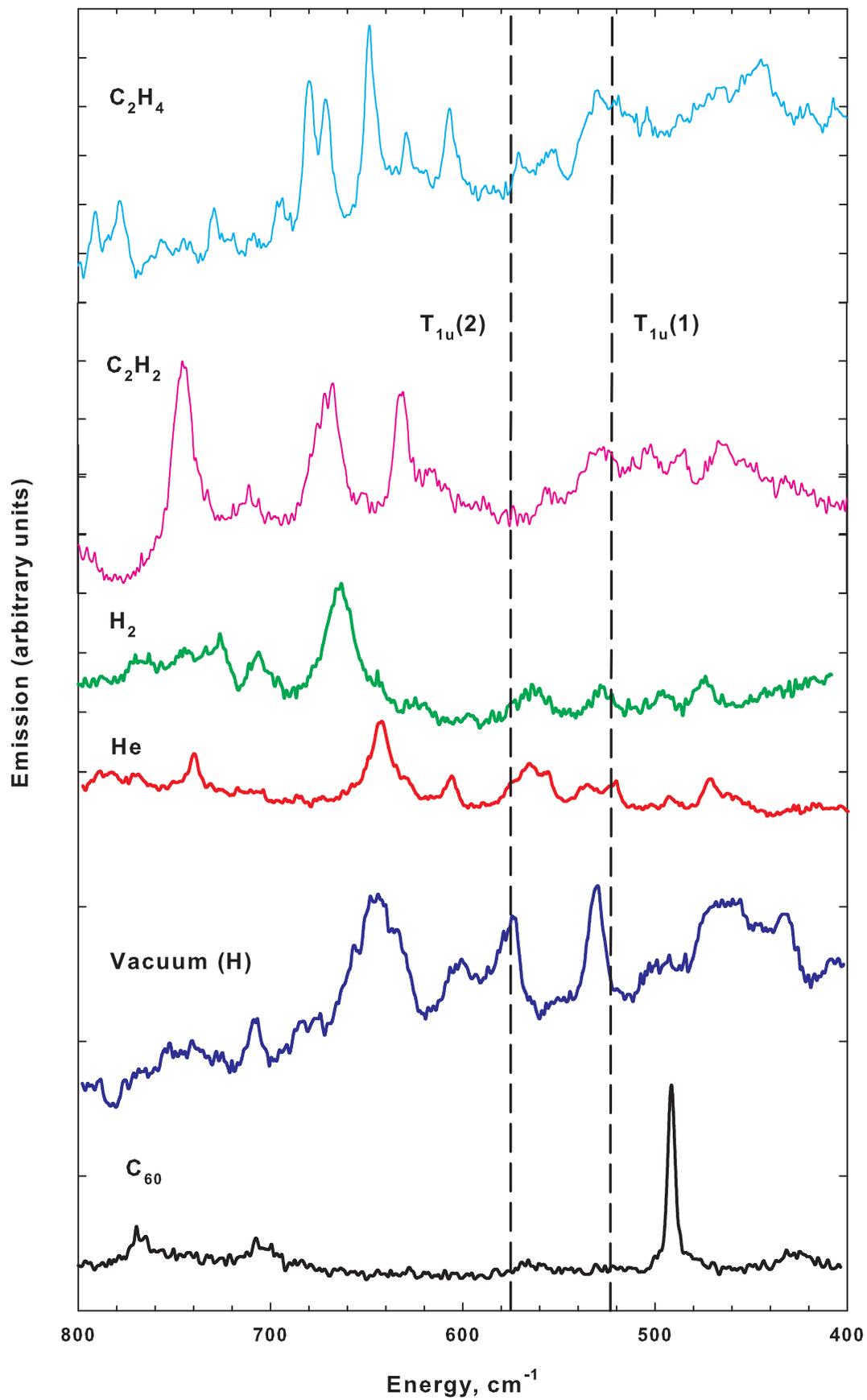

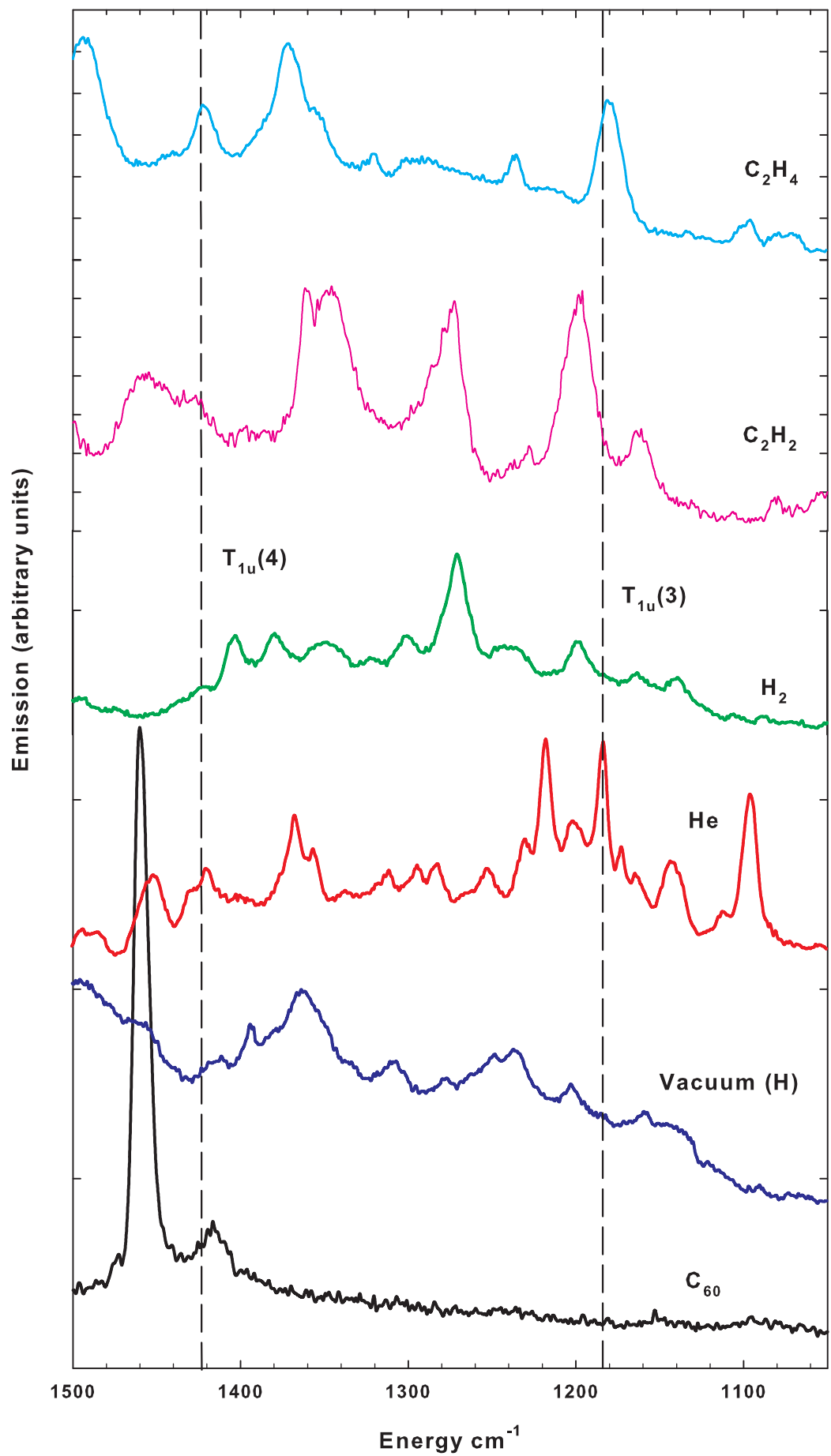